# W-RNN: News text classification based on a Weighted RNN

Dan Wang, Jibing Gong and Yaxi Song

*Abstract*—Most of the information is stored as text, so text mining is regarded as having high commercial potential. Aiming at the semantic constraint problem of classification methods based on sparse representation, we propose a weighted recurrent neural network (W-RNN), which can fully extract text serialization semantic information. For the problem that the feature high dimensionality and unclear semantic relationship in text data representation, we first utilize the word vector to represent the vocabulary in the text and use Recurrent Neural Network (RNN) to extract features of the serialized text data. The word vector is then automatically weighted and summed using the intermediate output of the word vector to form the text representation vector. Finally, the neural network is used for classification. W-RNN is verified on the news dataset and proves that W-RNN is superior to other four baseline methods in Precision, Recall, F1 and loss values, which is suitable for text classification.

## I. INTRODUCTION

On account of the certainty and comprehensibility of its expression, text has become the popular way of information expression and transmission. Text classification is an extremely important research direction [1]. Text classification has many practical application, these include news classification [2-4], emotional analysis and answering question system. We can use the text classification method automatically categorizes news data, afterward displays it on a news site for convenience people to glance over. In MicroBlog or Twitter, people can classify the words they post, and judge people's attitudes or opinions towards an event [5-6]. Specifically, comment information is taken as the text object to be classified, and it is divided into two categories, one is positive attitude and the other is negative attitude. According to the above settings, comment information can be classified. Event Categorization is also possible [7-8]. Similarly, the text classification model can be applied to the field of e-commerce. According to the evaluation of buyers on some commodities, the preferences of buyers are analyzed, so as to better respond to the needs consumers. In the field of Natural Language Processing (NLP), text classification can be seen as part of NLP. For example, users' questions can be classified, after that combined with other algorithms to obtain more accurate answers in the Question Answering System (QAS). In a word, text classification is widely used in real life, which has great economic and social value for accurate text classification.

The current challenges of text classification are as follows: 1) most of the work simplifies the text into Bag Of Words (BOW), with the semantic unit of words as the feature item, while ignoring the relationship between semantic units. 2) the general dimension of document representation is high, resulting in semantic sparseness, and 3) most algorithms cannot solve the problem of time series change and long-term dependence.

To address these issues, the main contributions of this study lie in the following three aspects:
1) Word2vec replaces the commonly used BOW, reducing the document representation dimension and effectively alleviating the semantic sparse problem;
2) Weighting the context of the word2vec by obtaining the intermediate vector of the words through the RNN in units of LSTM, thereby representing the document vector;
3) Introduce the W-RNN text classification process in detail, classify the above document representation method with neural network, afterward, verify the effectiveness of W-RNN method through experiments, and analyze the performance of related algorithms.

Traditional text feature vectors have problems such as large dimensions, independent vocabulary, and inability to express textual semantic information. In this study, we proposed method weighted recurrent neural network (W-RNN) use word2vec to represent vocabulary, and then word vectors are used to represent text to extract text semantic information effectively. Meanwhile, according to the characteristics of text data serialization, the text representation method of RNN automatically extracting text features is improved. The intermediate output vector of each word in the RNN structure is automatically weighted and summed as a document vector, which is classified by using neural network. W-RNN can comprehensively synthesize the semantic information of the whole text, and carry out deeper feature extraction on the text, effectively improving the text classification effect.

The rest of our paper is organized as follows: Section 1 is the introduction. Section 2 describes the theoretical basis of the W-RNN method proposed in this paper. Section 3 introduces the experiment and results in detail. In section 4 describes the research progress of text classification. Section 5 is the conclusion.

Manuscript received October 9, *. This work was supported by the Hebei Natural Science Foundation of China (Grant No. F2015203280) and the Graduate Education Teaching Reform Project of Yanshan University (Grant No. JG201616).

The authors are with the School of Information Science and Engineering, Yanshan University, Qinhuangdao 066044, China (e-mail: wangdanysu8100@163.com, gongjibing@163.com ,18712798790@163.com).

## II. OUR APPROACH: W-RNN

### A. Framework

The W-RNN method proposed in this study is shown in Fig.1, which consists of three parts, Long Short-Term Memory (LSTM), WD-vector document representation, and NN classification. The LSTM obtains the vocabulary intermediate feature vector (*WM-vector*) of word2vec, and the WD-vector document representation indicates that the weighted *WM-vector* is summed as the text representation vector that trained as the input of the neural network classification to obtain the final classification result.

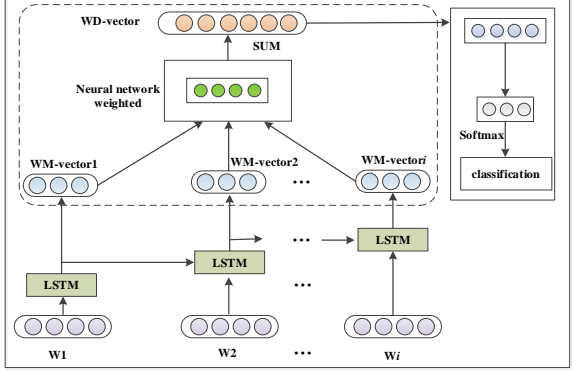

Fig. 1. Framework of W-RNN

The lower part of the Fig. 1 means to LSTM, which serializes the text data $W_t$ into the LSTM unit for better memory and access. The *WM-vector* of each word is obtained through the processing of three gate structures. The dotted box is the second part, whose parameters automatically obtained by the neural network are used as the weights of the *WM-vector*. The word vectors in each text are cumulatively summed by weights to get the representation vector of the document (*WD-vector*). The right part of the figure1 signifies the *WD-vector* training through the hidden layer structure of the neural network, after that the *Softmax* layer is optimized to obtain the classification result.

### B. Implementation of W-RNN

*a) Text length selection*

To extract the text serialization information using the RNN to achieve classification, we must first determine the length of the text. The length of the input text affects the quality of the RNN to a certain extent. So we firstly counts the length of each text in the dataset (the number of words appearing), and selects a suitable text length (sequence-length, *SL*), which is calculated by (1).

$$SL = \min(DLs)$$
$$s.\,t.\ \theta \leq \frac{DNum(DL < SL)}{DNum} \quad (1)$$

We need to set a threshold $\theta$ ($\theta \in (0,1)$), *DNum* represents the number of all documents, *DNum* (*DL < SL*) means to the number of documents whose text length is smaller than *SL*, and *DLs* represents the set of all document lengths appearing in the dataset. The smallest *SL* that can satisfy the constraint is found as the fixed length of the text in the actual process. In this paper, the threshold is set to 0.85 according to experience. If the number of documents whose length is less than or equal to *SL* in the corpus exceeds $\theta$, and *SL* is the minimum value that satisfies the condition, then this value is a fixed length of text.

After determining the *SL* in the dataset by the above document length selection method, the *SL* is truncated (that is, intercept words of fixed length), and the text having the length smaller than *SL* is directly filled with zero. This ensures that most of the information in the document is preserved and contains less redundant information. Such input simplifies the training process of the RNN.

*b) LSTM*

In the text classification, LSTM unit is used to extract the text vocabulary and pre-text information. Multiple LSTM are horizontally superimposed to form RNN to extract the text serialization information, afterward, text classification is performed. The process is illustrated in Fig. 2.

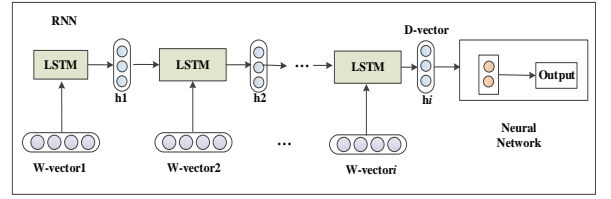

Fig. 1. Basic RNN

In Fig.2, *W-vector₁*, *W-vector₂*, and *W-vectorᵢ* are word vectors in the text, which are represented by word2vec and input them to each LSTM according to the order in which the words appear in the text (indicated by LSTM box in the figure). Each word2vec generates a hidden state (intermediate feature vector of the words, *h1* and *h2* in the figure) h through a single or multiple LSTM. The *h* as part of the input of the next LSTM, cycle training in the above manner, and using the last output vector in RNN as the document vector (*D-vector*), then classifying the *D-vector* using a neural network.

The calculation process of the LSTM is realized by three gate structures, namely, forget gate, input gate and output gate, retaining and filtering text information through these three gates.

Forget gate can calculate how much text information will be discarded before the current time, its calculation process is illustrated in (2).

$$f_t = \sigma\left(W_f\left[h_{t-1}, x_t\right] + b_f\right) \quad (2)$$

$W_f$, $b_f$, and $\sigma$ are the weights, bias and Sigmoid activation functions in the neural network respectively. Taking the former information and the word vector of the current vocabulary as input, through the neural network training, how much semantic information that has appeared in the document will be forgotten, and the forgetting information has no positive effect on the subsequent training process.

The input gate determines how much of the currently entered data will be used, including two parts, one that determines which information needs to be updated, and another that produces the content to be updated. In the text classification, it is also determined whether the current vocabulary and the previous information have a positive effect on the document representation.

The updated partial calculation process is shown in (3).

$$g_t = \sigma\left(W_g\left[h_{t-1}, x_t\right] + b_g\right) \quad (3)$$

The calculation process of candidate update content is presented in (4).

$$c_t = \sigma(W_c[h_{t-1}, x_t] + b_c) \quad c_t = \sigma(W_c[h_{t-1}, x_t] + b_c) \quad (4)$$

Where $W_g$ and $W_c$ are weights, $b_g$ and $b_c$ are bias, and $\sigma$ is the *Sigmoid* activation function.

LSTM updates the state of the cell through input and forget gates as the (5) follows.

$$s_t = f_t \times s_{t-1} + g_t \times c_t \quad (5)$$

The output gate determines the output of the LSTM. Each word vector and the previous information will have an output gate to control the content of the output information. The output gate is shown in (6), where the activation function is *Sigmoid*, and $W_o$ is the weight, $b_o$ is the bias.

$$o_t = \sigma(W_o[h_{t-1}, x_t] + b_o) \quad (6)$$

The final output of the LSTM is the current cell state controlled by the output gate. Our article considers the output of each LSTM as the intermediate feature $h_t$ (the intermediate feature of each word) integrated with each word and the previous information, useful pre-information in the document is transmitted through $h_t$, the calculation is shown in (7).

$$h_t = o_t * \tanh(s_t) \quad (7)$$

$h_t$ can capture the information of the current input vocabulary and the previous vocabulary well, which is the combination of the pre-text and the general semantic information of each vocabulary in the text.

The document is composed of a series of sequential words that is serialized information covering the semantics of the text. High-quality feature extraction of the context of the vocabulary before and after can effectively improve the accuracy of text classification. In LSTM, each vocabulary in the text serialization data is merged with the $h_{t-1}$ calculated by the previous LSTM through three different neural network gate structures, and is continuously transmitted so that the deeper semantic information of the text can be effectively preserved.

*c) W-RNN*

W-RNN changes the document vector based on the RNN that the serialized document vector does not simply rely on the output of the last LSTM. Instead, the output of each word in the RNN at each moment is used as the word middle vector (*WM-vector*). Let the length of the document be *SL*. First, train the *WM-vector* of each word through the *SL* neurons, after that weight each *WM-vector* according to the weight of the neural network training, and sum the weighted *WM-vector*. A weighted document vector (*WD-vector*) is formed and classified by a neural network. As a final step, the probability distribution of the category to which document belongs is obtained, therefore the text category is further known. The specific implementation process of automatic weight learning is shown in Fig. 3. The W-RNN principle is shown in Algorithm 1.

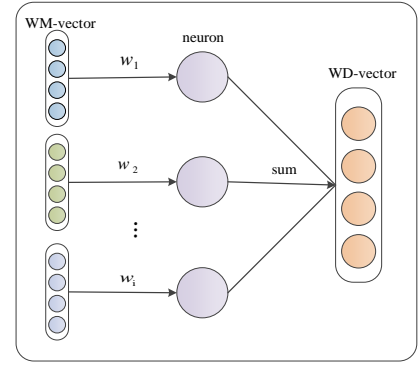

Fig. 2. Weight learning

In Fig. 3, *WM-vector* is a word middle vector, $w_1$, $w_2$, etc. are weights, and the *WM-vector* are weighted and summed to form the *WD-vector*. It is shown in (8).

$$WD\text{-}vector = \sum_{i=1}^{seq\_length} w_i * WM\text{-}vector_i \quad (8)$$

*WM-vector$_i$* means to the intermediate feature output vector of the *i-th* word, which is the weight by neural network. *WD-vector* signifies the text representation vector after the weighted and summation of the *WM-vector$_i$*.

| Algorithm1: W-RNN |
|---|
| **Input**: Dataset(A total of C categories and N samples) |
| **Output**: Category of each news data: Y |
| **Step**: |
| 1. Initialize WD-vector[N]; |
| 2. **for** doc in Dataset: |
| 3.   **for** word in doc: |
| 4.     get document length; |
| 5.   **end for** |
| 6. **end for** |
| 7. select SL; |
| 8. train word vector; |
| 9. RNN = new SL * LSTM (); |
| 10. **for** word in document: |
| 11.   WD-vectori =RNN. LSTM(word2vec); |
| 12. **end for** |
| 13. Wi = neural Network(SL); |
| 14. WD-vector[i]=SUM(Wi* WD-vectori); |
| 15. Y = neural classification(WD-vector); |

Although LSTM can better learn the textual information and selectively record the semantic information, it still loses some valid thing during the iterative training, and cannot completely guarantee that the effective information is recorded. If only the last $h_t$ in the RNN is used as the text representation vector, all the information of the document cannot be fully extracted. W-RNN pays more attention to important vocabulary information that has positive effects on classification, and reduces attention to unimportant word. Weighting can make significant words in the document have higher weight, so the influence of these words on the text classification is greater. Weighting *WM-vector* can capture the central semantics of the document, and further extract the semantic information from total paragraph, as shown in Fig. 4. It is also an innovation in text representation that improves the accuracy of classification.

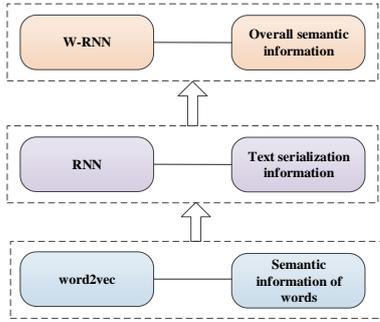

Fig. 3. W-RNN information extraction process

## III. EXPERIMENTS

### A. Experimental Settings

To evaluate our method, we developed W-RNN in 20 News Groups Dataset that has 20 categories and the total sample size is 18820. Many of the categories fall into overlapping topics; for example 5 of them are about company discussion groups and 3 of them discuss religion. Other topics include in News Groups are: politics, sports and sciences. The number of documents contained in each category is around 1000, and the data of each category is relatively balanced in general. The length of words in each document in 20 News Groups Dataset is counted quantitatively, and the results are shown in Fig. 5.

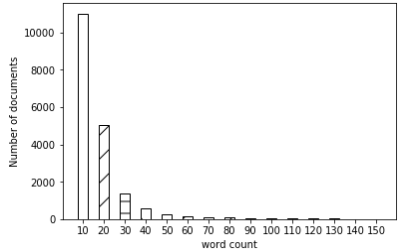

Fig. 5. Document length statistics for 20 News Groups Dataset

In Fig. 5, the x-coy-coordinate signifies the number of words from 100-1500 (10-150 in the figure), and the y-coordinate is the number of texts. Depending on statistics, the majority of the text length in the data set is within 300. According to the text length selection method, select 300 as the fixed text length.

We train all our model on a Nvidia GeForce GTX1080Ti GPU card with 11GB RAM. We divide the datasets for evaluating our model: 90% training set and 10% test set. The dimension size of word2vec is set to 200 in practice and the length of the vocabulary is 40439. Our proposed algorithm is consisted of a hidden layer with 128 units. We use the cross-entropy loss function and the activation function is *Relu* to training, the number of convolution kernel is 128 in W-RNN, minibatch size is set to 128 while the learning rate is set to 0.01. To avoid over-fitting, we employ regularization rule and set λ of regularization rate to 0.01, we update the parameter using *ADAM optimizer*. Precision (P), Recall (R) and F1 are used as evaluate metrics.

### B. Baseline methods

Here, we explain how our architecture W-RNN outperforms existing W-RNN. To conduct our comparison, we selected four baselines:

**DNN**: Deep neural network, with the word embedding data of the text as input, trains the input text with a multi-layer simple neural network, and finally obtains the text classification result.

**RNN**: Recurrent neural network, training through LSTM, taking text word2vec as input, using word2vec to represent words, classifying the final output of sequence in RNN with neural network, and finally obtaining text classification result [9].

**Bi-RNN**: Bidirectional recurrent neural network, training through two-direction LSTM, using word2vec to represent text and as input, combining the final output results in two directions, and next using simple neural network training to obtain classification results [10].

**CRNN**: The result of the CNN is used as the input of the RNN. Through the LSTM training, the last output of the text sequence is input into the simple neural network model, and finally the classification result is obtained.

### C. Performance and comparison

Randomly selected 10% of the data was as the test set. The following experimental results are obtained through the above experimental settings, which are introduced below.

When DNN has the best effect on the test set, the accuracy rate is close to 56%, and the loss value is about 1.3. The accuracy rate on the training set is about 59%, and the loss function value on the training set is about 1.2.

The performance of RNN is shown in Fig. 6, whose x-coordinate means to the number of iterations, the y-coordinate is the precision and the value of loss function.

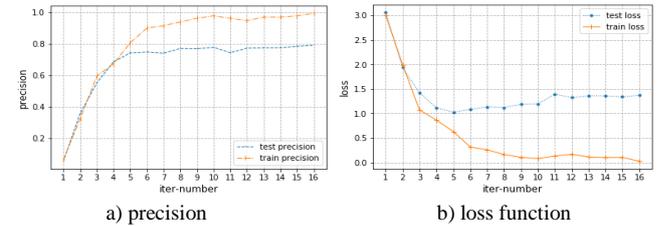

a) precision        b) loss function

Fig. 6. RNN's result

It can be observed in Fig. 6 that after the fifth iteration, the accuracy of text classification on the training set and test set both tend to be stable. At this time, the value of loss function does not change much. When the effect of the RNN is optimal, the accuracy rate on the test set is about 78.91%, and the loss function value is 1.36. The accuracy on the training set is 99.21%, and loss function is 0.027.

Fig. 7 shows the result of Bi-RNN, whose x-coordinate signifies the number of iterations, the y-coordinate is the precision and the value of loss function.

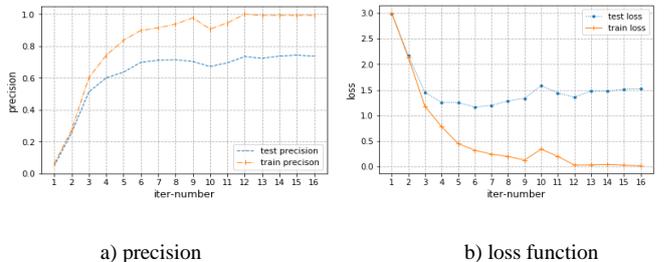

a) precision        b) loss function

Fig. 7. Bi-RNN's result

As can be seen from Fig. 7, Bi-RNN gradually stabilizes after iteration 6 times. The highest accuracy rate is 74.31% on the test set, the lowest loss function value is 1.24, and the accuracy on the training set is 99.22%. The value of the loss function is 0.029.

CRNN results are presented in Fig. 8.

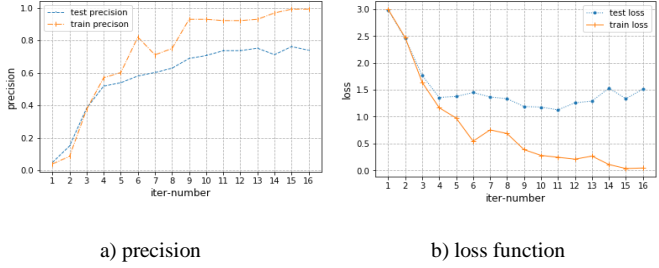

a) precision    b) loss function

Fig. 8. CRNN's result

Fig. 8 demonstrate that the model tends to be stable after 10 iterations in the process of CRNN training, and the accuracy rate on the test set reaches 76.16%, the loss function value on the test set is 1.33. However, the accuracy rate fluctuates greatly in the model training process.

The experimental results of W-RNN are shown in Fig. 9.

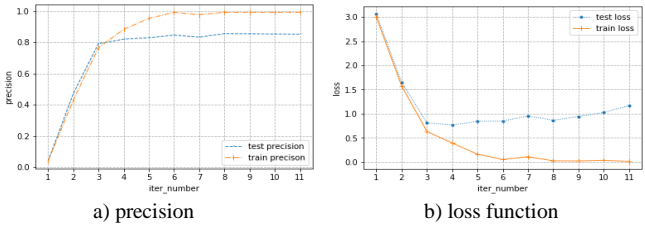

a) precision    b) loss function
Fig. 9. CRNN's result

From Fig. 9, the W-RNN classification effect is better than other baseline methods. After iterating 5 times, the model tends to be stable. When the effect is best, the accuracy rate on the test set is about 85.55%, and the loss function value is 0.86. The accuracy rate on the training set is 99.2% and the loss function is 0.056.

Table 1 and Fig. 10 are obtained by comparing the experimental results of text classification based on RNN.

TABLE I: COMPARISON OF EXPERIMENTAL RESULTS BASED ON RNN TEXT CLASSIFICATION

|  | DNN | RNN | Bi-RNN | CRNN | W-RNN |
|---|---|---|---|---|---|
| Precision | 56% | 79% | 75% | 78% | 85% |
| Recall | 55% | 77% | 74% | 76% | 84% |
| F1 | 53% | 78% | 75% | 77% | 84% |
| Loss value | 1.2 | 1.36 | 1.5 | 1.33 | 0.86 |

The text is classified by different neural networks. The W-RNN is higher on Precision, Recall and F1, and the loss function value is also small. This shows that when the W-RNN classifies, the intermediate feature output vector of each word in the RNN is weighted to form the *WM-vector*, which can handle sequences of arbitrary length and obtain long-term dependencies, effectively combining the upper and lower associations between words, it also can extract the deep meaning of the text to improve the accuracy of text classification. In addition, LSTM takes the state of the last hidden layer as an output and then classifies it with the *Softmax*; our method is the result of saving each hidden layer of the LSTM, making full use of it and combining the output with the corresponding word vector.

Bi-RNN classification accuracy rate is not as high as RNN. It may be due to the information in bi-direction of the text is comprehensively used, some information are cancelled, or the result is deviated owing to certain noise and splicing errors in the dataset. CRNN directly extracts the local feature of the input vector of the word embedding, and compresses the text data first. This process destroys the serialization feature of the text. When the RNN is used, the semantic information of the text cannot be extracted well, so the classification effect is not good.

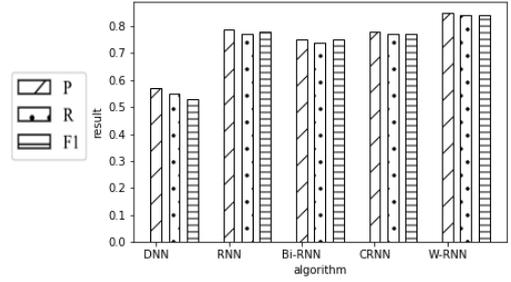

Fig. 10. Compare results

## IV. RELATED WORK

The references [11-13] have a related introduction to the text representation method of TF-IDF. However, TF-IDF based on statistical information has some defects: the dimension of the text representation is large, and each text is represented as a vector of the length of the vocabulary, which has extremely high requirements on storage space and computing performance. Besides, the text vector is sparse, which affects the accuracy of the classification. The words in TF and TF-IDF are independent of each other, and do not consider the relationship between vocabulary, these include synonyms, synonyms and antonyms, etc., and lack of expression of semantic information of the text. Another vocabulary representation is word2vec [14]. Through word embedding [15-16] (CBOW or Skip-Gram[17]), the word vector of each word in the corpus can be learned, after that the weight is shared by the neural network and calculated the vector representation of each word. Word2vec solves the problem of One-hot representation dimension, while the word2vec can learn the deep semantic information of vocabulary [18-19], and better extraction of words semantic information is of great significance to improve the accuracy of text classification. The emergence of word2vec changes the traditional way of using text statistical features to represent text. It can well represent the relationship between words and well reflect the semantic information of words. Meanwhile, the dimension of the word vector can be specified, so there is no problem of data sparse. Knowledge regularized word representation [20] incorporate prior knowledge for learning distributed word representations. Documents can also be represented through topics model[21].

There is no need for artificial feature selection in neural network classification, and it has powerful learning ability, which allows the model to automatically learn certain content and patterns. The gradient descent is used to reduce the loss function, back propagation, and various parameters (connection weights) in the neural network are adjusted [22]. Text is serialized data, which can be classified by RNN. The unit structure in RNN is reused, and the serialized data of the current moment and the hidden state data at the previous moment are iteratively input into the unit structure, and then classified by neural network. Each neuron has a corresponding activation function. The commonly used activation functions are *Tanh*, *Sigmoid*, and *ReLU* [23].

The simple neural network text classification uses a multi-layer neural network to classify texts. Taking the reference [24] as an example, it can classify various data and continuously learn weights through forward and backward propagation to reduce the error between predicted and true values. The simple neural network's structure design is not targeted to the text classification, so the serialization information of the text cannot be well extracted, and the network structure setting is relatively simple, which is not suitable for the text classification. The Bi-directional RNN [25] (Bi-RNN) is based on the one-way RNN structure, and simultaneously trains the text in both positive and negative order, synthesizes the training results in both directions of the text, then classifies the text with bi-directional text features. CRNN[26] combines CNN and RNN to solve classification problem, firstly use CNN for training, then secondly calculate the result of CNN as input of RNN. As a final step, the RNN training results are classified. The pooling layer effectively reduces the amount of data, reduces feature dimensions, and afterward, captures the sequence information of text content through LSTM to obtain long dependencies in the text. However, reducing the dimension of the text extraction feature may cause damage to the text semantic information, and the context information of fuzzy text is not suitable for text classification.

## V. CONCLUSION

In this study, the various process of text classification is studied in detail. In the aspect of text representation, through analyzing the process of automatically extracting text features from RNN, a W-RNN text classification algorithm is therefore proposed to improve RNN. The vector of text representation is achieved by weighting the intermediate features of text vocabulary, after that the text is categorized by using neural network. The specific process is as follows:1) Utilize word vector represents the text vocabulary, and the RNN is used to obtain the intermediate feature vector of each word, 2) Automatically train the weight, and the weight sum of the intermediate feature vector of each word as the text representation vector WD-vector. 3) A simple neural network is utilized to classify text. We implemented W-RNN with 20 News Groups Dataset, which has 20 categories, and the number of news documents in each category is evenly distributed. By comparing with four baseline methods, it is concluded that the W-RNN effect is better. After five iterations, the model tends to be stable. When the model has the best effect, the accuracy rate on the verification set is about 85.55%, the loss function value is 0.86, and the accuracy rate on the training set is 99.2%, the loss function is 0.056. It is shown that the W-RNN can form the text vector by weighting the intermediate feature output vector of each word in the RNN, which can extract the deep paragraph information of the text and improve the accuracy of the text classification.

The W-RNN text classification method improves the accuracy of text classification to a certain extent, but it still has some shortcomings and can be improved. In the process of neural network training, how to effectively adjust the parameters in each algorithm has yet to be studied.

## REFERENCES

[1] Vandana Korde, C Namrata Mahender. Text classification and classifiers: A survey. *International Journal of Artificial Intelligence & Applications*. vol.2, no.3, 2012.
[2] Peng Hao, Li Jianxin, He Yu, et al. Large-scale hierarchical text classification with recursively regularized deep graph-cnn. *Proceedings of the 2018 World Wide Web Conference*, pp.1063-1072, 2018.
[3] He Yu, Li Jianxin, Song Yangqiu, et al. Time-evolving text classification with deep neural networks. *Proceedings of the 27th International Joint Conference on Artificial Intelligence*, pp.2241-2247, 2018.
[4] Arif Muhammad Hassan, Li Jianxin, Iqbal Muhammad, et al. Optimizing XCSR for text classification. *2017 IEEE Symposium on Service-Oriented System Engineering (SOSE)*. pp.86-95, 2017.
[5] Kevin Stowe, Jennings Anderson, Martha Palmer. Improving classification of Twitter behavior during hurricane events. *Proceedings of the Sixth International Workshop on Natural Language Processing for Social Media*, vol.1, no.1, pp.67-75, 2018.
[6] Li Ma, Shasha Li. Research on text classification based on word embedding. *Computer & Digital Engineering*, vol. 2, no.47, pp. 281-303, 2019.
[7] Peng Hao, Li Jianxin and Gong Qiran, et al. Fine-grained Event Categorization with Heterogeneous Graph Convolutional Networks. *Proceedings of the 28th International Joint Conference on Artificial Intelligence*, pp.3238-3245, 2019.
[8] Bao Mengjiao, Li Jianxin, Zhang Jian, et al. Learning Semantic Coherence for Machine Generated Spam Text Detection. 2019 International Joint Conference on Neural Networks (IJCNN). 2019.
[9] Yu Qiao, Shuwei Yao. Research on different text representation methods of large-scale protein function prediction. *Microcomputer Applications*, vol.7, no.34, pp.1-5, 2018.
[10] Zhiqiang He, Jian Yang, and Changling Luo. Combination characteristics based on BiLSTM for short text classification. *Intelligent Computer and Applications*, vol.2, no.9, pp. 21-27, 2019.
[11] Abdisa Demissie Amensisa, Seema Patil. A survey on text document categorization using enhanced sentence vector space model and bi-gram text representation model based on novel fusion techniques. *2018 2nd International Conference on Inventive Systems and Control (ICISC)*, 2018.
[12] Ammar Ismael Kadhim, Yu-N Cheah. Text Document Preprocessing and Dimension Reduction Techniques for Text Document Clustering. *2014 4th International Conference on Artificial Intelligence with Applications in Engineering and Technology*, 2014.
[13] Thabit Sabbah, Ali Selamat, Md. Hafiz Selamata, et al. Hybridized term-weighting method for Dark Web classification. *Neurocomputing*, vol.3, no.173, pp. 1908-1926, 2016.
[14] Yoshua Bengio, Réjean Ducharme, Pascal Vincent. A neural probabilistic language model. *Journal of Machine Learning Research*, vol.1, no.3, pp. 1137-1155, 2003.
[15] Yikang Shen, Wenge Rong, Nan Jiang, Baolin Peng, Jie Tang, Zhang Xiong. Word embedding based correlation model for question/answer matching. *In Proceedings of the 31st AAAI Conference on Artificial Intelligence (AAAI'17)*, pp. 3511-3517, 2017.
[16] Tiansi Dong, Chrisitan Bauckhage, Hailong Jin, Juanzi Li, Olaf Cremers, Daniel Speicher, Armin B. Cremers, Joerg Zimmermann, Imposing category trees onto word-embeddings using a geometric construction. *ICLR 2019*, 2019.
[17] Liner Yang, Xinxiong Chen, Zhiyuan Liu. Improving word representations with document labels. *IEEE/ACM Transactions on Audio, Speech, and Language Processing (ACM TASLP)*, vol. 25, no.4, pp.863-870, 2017.
[18] Tuba Parlar, et al. QER: a new feature selection method for sentiment analysis. *Human-centric Computing and Information Sciences*, 2018.
[19] Berna Altınel, Murat Can Gani. A new hybrid semi-supervised algorithm for text classification with class-based semantics. *Knowledge-Base d Systems*, vol.108, no.1, pp.50-64, 2016.
[20] Yan Wang, Zhiyuan Liu, Maosong Sun. Incorporating linguistic knowledge for learning distributed word representations. *PLoS ONE* vol.10, no.4, 2015: https://doi.org/10.1371/journal.pone.0118437.
[21] Lei Hou, Juan zi Li, Xiao li Li, et al. Learning to align comments to news topics. *ACM Transactions on Information Systems*, 31pages, 2017.
[22] Jinhong Li. TensorFlow entry principle and advanced practice for deep learning. China Machine Press, 2018, pp.111-114.
[23] Ming Tu, Xiang Liu, and Shuchun Liu. Python natural language processing practice. China Machine Press, 2018, pp.211-212.

[24] FrancescoGargiulo, StefanoSilvestri, et al. Deep neural network for hierarchical extreme multi-label text classification. *Applied Soft Computing*. vol.79, pp.125-138, 2019.

[25] Gang Liu, Jiabao Guo. Bidirectional LSTM with attention mechanism and convolutional layer for text classification. *Neurocomputing*. vol.337, pp.325-338, 2019.

[26] Lei Chen, Shaobin Li. Improvement research and application of text recognition algorithm based on CRNN. *SPML '18 Proceedings of the 2018 International Conference on Signal Processing and Machine Learning*. 2018.

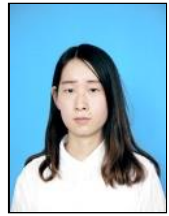
**Dan Wang** was born in Bao Ding, Hebei Province on 28 February 1993. She is a graduate student in the School of Information Science and Engineering, Yanshan University. She was admitted to bachelor's degree in computer science and technology in 2017. Her main research interests include machine learning and news mining.

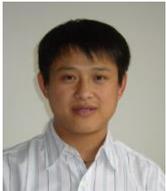
**Jibing Gong** received his PhD degree in Institute of Computing Technology, Chinese Academy of Sciences, China. He is a professor in the School of Information Science and Engineering, Yanshan University. He is head of the Knowledge Engineering Group (KEG) research team in Yanshan University. His main research interests include big data analytics, heterogeneous information network, machine learning and data fusion.

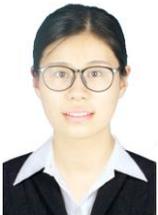
**Yaxi Song** was born in Bao Ding, Hebei Province on 17 November 1994. She is a graduate student in the School of Information Science and Engineering, Yanshan University. She was admitted to bachelor's degree in computer science and technology in 2017. Her main research interests include machine learning and social networks.